\documentclass[preprint2]{aastex}
\slugcomment{submitted to {\it The Astronomical Journal}}
\begin{document}
\title{Photometric Abundances for G-Dwarfs: A Cautionary Tale}
\author{Bruce A. Twarog, Barbara J. Anthony-Twarog, and Delora Tanner}
\affil{Department of Physics and Astronomy, University of Kansas, 
Lawrence, KS 66045-2151}
\affil{Electronic mail: twarog@ukans.edu,bjat@ukans.edu,dtanner@ku.edu}
\begin{abstract}
Analysis of cluster and field star $uvby$ data demonstrates the
existence of a previously undetected discrepancy in a widely used
photometric metallicity calibration for G dwarfs. The discrepancy is systematic
and strongly color-dependent, reducing the estimated [Fe/H] for stars above 
[Fe/H] $\sim -0.2$ by between +0.1 and +0.4 dex, and creating a 
deficit of metal-rich stars among dwarfs of mid-G and later spectral type. The 
source of the problem, triggered for stars with $b-y$ greater than 
about 0.47, appears to be an enhanced metallicity dependence for 
the $c_1$ index that increases as temperature declines. The link 
between $c_1$, normally a surface gravity indicator, and metallicity 
produces two secondary effects. The deficit in the photometric abundance 
for a cool dwarf is partially compensated by
some degree of evolution off the main sequence and 
cool dwarfs with metallicities significantly above
the Hyades are found to have $c_1$ indices that classify them as giants.
The potential impact of the problem on stellar population studies is discussed.

\end{abstract}
\section{INTRODUCTION}
The {\it Hipparcos} \citep{PE97} era in 
stellar astronomy has produced a
revolution in our ability to define the evolutionary state of large numbers
of stars in a wide array of stellar classes and subclasses. The capability
exists to define reliably the distance and tangential velocity of thousands
of stars and place them, with a high degree of accuracy, at the appropriate
location in some form of color-magnitude diagram (CMD). The translation of the
CMD location into a point on an evolutionary track for a star of a given
mass requires information external to the {\it Hipparcos} survey, 
in particular, the conversion between the observational and the theoretical 
plane that depends, in part, on the metallicity of the star.

This quantum leap in the quality and quantity of data supplied by 
{\it Hipparcos} has generated a tremendous gap between analyses based on these
data as opposed to the more traditional, ground-based
observations. A prime example is presented by stellar 
spectroscopy. Though low-resolution spectroscopic catalogs such as 
the {\it Michigan Catalog} \citep{HC75,HO78,HO82,HS88,HS99} compare 
favorably to the space-based sample in number,
high-resolution, spectroscopic catalogs listing abundances, 
e.g. \citet{CA01}, contain only a few thousand stars of mixed range and
reliability. In the absence of a spectroscopic equivalent of {\it Hipparcos},
investigators have used a variety of approaches to fill the void,
the most common being analysis of photometric indices. Of the many photometric
systems available, arguably the most valuable and most used 
for large-scale studies is
the Str\"omgren $uvby$ system. The intermediate-band filters of the system have
been selected to disentangle the overlapping effects of varying temperature,
metallicity, and surface gravity and, over the last 35 years, the indices
have been calibrated to supply information on stars over the entire
range in temperature, luminosity, and abundance. Moreover, the
photometric sample numbers over 60,000 stars and 
is currently available on a uniform system via the efforts of \citet{HM98}. 
Though composite catalogs can often be of questionable value due 
to the inadequate nature of the transformations among different observers, 
the catalog of \citet{HM98}
is dominated by the extraordinary sample collected over the last 20 years
by \citet{OLS83,OLS84,OLS93,OLS94} and others \citep[e.g.,][]{SN88,SP93}, 
often using the same instrumentation and reduction procedures. Thus, 
the transformation corrections among
the various published catalogs are small to negligible.  

For stellar population investigations of the solar neighborhood, $uvby$
photometry has long been a mainstay, with an emphasis on stars of spectral
type A through early G. Because the $c_1$ index supplies a measure of $M_V$,
it permits a determination of the location of the star in the CMD and,
in combination with the metallicity via $m_1$ and temperature via $b-y$ or
H$\beta$, an age through comparison with appropriate isochrones
\citep{TW80, ME91}. With the availability of {\it Hipparcos} parallaxes to
determine $M_V$, the role of the $uvby$ system has shifted to one of supplying
internally consistent abundances, while the spectral range of interest has
moved redder, to stars of type late G and early K. A few of the recent
examples of the use of the system may be found in 
\citet{LA00, RP00, HA01, FE01, RE02}. As invariably happens when new data 
open up previously unexplored avenues
of research, the analyses lead to contradictions with earlier work and
with each other; examination of the citations noted above prove that the
$Hipparcos$ data analyses are no exception. In evaluating such contradictory
evidence, it is crucial to isolate the real effects of galactic and
stellar evolution from the artifacts of the technique. The purpose of this
paper is not to review the strengths and weaknesses of the many 
recent solar neighborhood investigations tied to $uvby$ photometry. 
Our purpose is to highlight an inherent problem in the
$uvby$ metallicity calibration for disk G and K stars, a calibration
that is being used with increasing frequency in such studies.

In Sec. 2, we will present the evidence that a problem does, in fact, exist. In
Sec. 3, we track down the source of the discrepancy, while 
investigating its role in population studies in Sec. 4. Sec. 5 provides a
summary of our conclusions.

\section{The Problem}

Over the last 15 years, we have been involved in the development of a
photometric system known as {\it Caby} photometry.
The {\it Caby} system represents an extension of the traditional four-color,
$uvby$ intermediate-band photometric system to a fifth filter centered on the
H and K lines of Ca II. Details of the filter definition and design as well as 
the fundamental standards may
be found in \citet{ATT91} while an extensive catalog of stars observed on the
system and tied to the $b-y$ scale of \citet{OLS93} may be found in 
\citet{TAT95}. The filter was designed initially with metal-deficient stars 
in mind, as demonstrated by numerous applications to date on normal field stars 
\citep{ATT91,AST92,ATT98,ATT00}, clusters \citep{ATC95,RE00}, and variables
\citep{BD96,HI98}.
Metallicity calibrations have been produced for both the metal-deficient giants
\citep{ATT98} and metal-deficient 
dwarfs \citep{ATT00}, but preliminary analysis indicated that
for dwarfs hotter than the sun, the $hk$ index, defined as {\it (Ca-b)-(b-y)},
remains metallicity sensitive for stars of solar abundance or higher 
\citep{TAT95}, a result consistent with the theoretical models of \citet{SO93}.
Because of the high metallicity of the Hyades relative to the typical star in
the field of the solar neighborhood, observations of this cluster
were obtained along with the field stars as a means of testing this
prediction.

In the course of analyzing a sample of over 100 Hyades dwarfs extending
to spectral type of late K, it was noted that the cooler dwarfs in the
Hyades occupied an unexpected location in the {\it hk, b-y} diagram. 
The data imply that higher metallicity shifts
the Hyades relation to $lower$ $hk$ at a given $b-y$ relative to the average
disk star, i.e., the Hyades stars formed an $upper$ bound in the figure rather
than a lower bound \citep{ATTY}. As a test of the metallicity option, the
stars located above the Hyades in the two-color relation were
identified and a check made for any published abundance information, 
photometric or spectroscopic. In the course of this search, it became
apparent that a serious deficiency of stars with solar metallicity or
higher exists among the cool stars in the published $uvby$ catalogs. The
source of this problem emerged when we attempted a differential comparison
of the photometric indices of the stars directly with those of the 
Hyades dwarfs of the $uvbyCa$ system.

To illustrate the problem, we make use of the $uvby$ data for two clusters,
Hyades and Praesepe, with abundances above solar as measured via spectroscopy
\citep[e.g.,][]{BO89,FB92} and photometry \citep[e.g.,][]{NI88,TAT97}).
The $uvby$ photometric data
for the Hyades come from \citet{OLS93,OLS94} while those of Praesepe come 
from \citet{REG91}. Comparison of 16 stars common to \citet{OLS93,OLS94} and
\citet{RF92} for the Hyades produced modest offsets between the two
photometric systems. The average differences among the various indices, in
the sense (OLS - REG), are +0.0199 $\pm$ 0.0259, +0.0079 $\pm$ 0.0061, 
--0.0078 $\pm$ 0.0060, and +0.0135 $\pm$ 0.0110 for $V$, $b-y$, $m_1$, and
$c_1$, respectively. Errors quoted are standard errors of the mean. Because
there is no direct means of comparing the Praesepe data of \citep{REG91}
to that of \citet{OLS93,OLS94}, we will adopt the offsets from the Hyades
comparison as representative of the Praesepe photometry.  
The cluster data were then adjusted to the system of \citet{SN88} 
using the transformation relations derived by
\citet{OLS93}. It should be emphasized in all cases over the color 
range of interest, $b-y$ between 0.39 and 0.59, the corrections are small and 
cannot be the source of the discrepancy. The cluster data are then processed
through the metallicity calibration for G dwarfs as derived in \citet{SN89},
the calibration currently adopted in most photometric analyses. The
resulting abundances, [Fe/H], are plotted in Fig. 1 as a function of $b-y$
for the Hyades (filled circles) and Praesepe (open circles).
Observations of the same Hyades star from two different catalogs were treated
as independent data and plotted as separate points.

For stars hotter than $b-y$ = 0.47, most stars scatter between [Fe/H] = 0 and
+0.2, indicative of the expected supersolar metallicity for stars in these
clusters. A more detailed analysis shows that there is 
weak evidence for a smooth
color dependence in the abundances, starting at [Fe/H] = 0.0 for
the bluest stars, the cooler F dwarfs, and rising above +0.13 at $b-y$ = 0.46.
For stars redder than $b-y$ = 0.47, there is a systematic decline
in the calculated [Fe/H], reaching a minimum between [Fe/H] = --0.2 and --0.3
at $b-y$ = 0.52 to 0.53, then rising rapidly toward +0.2 for stars at the cool
limit of the calibration. The fact that the pattern is the same for two clusters
of the same [Fe/H] observed independently by two different groups ensures that
this is not a problem tied to a systematic error in the photometry.

Though the photometry itself may not be the issue, perhaps the clusters, both
metal-rich and moderately young, suffer from some form of anomaly which
distinguishes them from the typical field star of the same [Fe/H] and 
temperature. An example of such an effect which immediately comes to mind
is the aptly named {\it Hyades anomaly} \citep[e.g.,][]{NI88,DO90,RF92,JT95}. 
To test this possibility, we
have processed the two primary, G-star $uvby$ catalogs \citep{OLS93,OLS94} 
through the same metallicity calibration, after applying adjustments for 
offsets relative to \citet{SN88}. The catalogs are treated
separately because they emphasize different temperature ranges and
are constructed using different selection criteria. 
The metallicity distribution as a function
of color is illustrated for \citet{OLS93} in Fig. 2a and for \citet{OLS94} in
Fig. 2b. Though the catalogs are not composed of a random sample of the cooler
stars in the solar neighborhood due to biases imposed by the spectral
classification selection and apparent magnitude limits, there is a broad
range in [Fe/H] at all colors. The striking element common to both samples
is the decline in the number of stars with [Fe/H] greater than 0.0 for
$b-y$ greater than 0.49. The only difference of note in comparison
with the cluster data is that the decline in the distribution
is triggered at a marginally redder color than found with the clusters, 
a point we will return to in Sec. 4.

Though it appears that a problem exists within the G-star calibration,
its source is less obvious. We emphasize that by the source of
the discrepancy, we are not referring to the explicit reason why the 
error was included in the original calibration. It is a trivial point 
to assume that when a calibration of this type exhibits discrepancies, 
the origin of the error can be traced back to some combination of 
photometric errors, spectroscopic abundance errors, or a sample size 
inadequate to cover the parameter space
of interest. The sample used by \citet{SN89} is no exception in that,
despite the explicit statement that the abundance calibrations were
appropriate between $b-y$ = 0.38 and 0.59 up to [Fe/H] = +0.4,
the sample is heavily weighted toward stars of halo metallicity.   

By source of the discrepancy, we instead refer to the term(s) in the
calibration relation that are inadequately or inappropriately
sensitive to the changes in the stellar parameters, producing
abundances that are not representative of the star. The rationale
for this search is that it isn't enough to simply state that the abundances
for some stars will be in error. If the errors are coupled to a specific
photometric index, the size of the deviation may be coupled to
a fundamental property of the star and lead to significant biases
within samples chosen on the basis of photometry. Since the G-star
calibration is being applied to extended samples for stellar
populations studies, as discussed in Sec. 4, it is crucial
to know which indices are primarily responsible for the discrepancy.

In the traditional approach
to metallicity calibration with the $uvby$ system, a metallicity index is
derived by determining the difference between the observed $m_1$ index of a
star and the value on a standard relation at the same temperature
as defined by a color index such as $b-y$ or H$\beta$ \citep{CR75,OLS84}.
The value of $\delta$$m_1$ is then transformed to [Fe/H] using some functional
form which includes the potential secondary effects of temperature 
and surface gravity variation. \citet{SN89} use the actual
indices, $b-y$, $m_1$, and $c_1$, rather than differential terms, and an
extensive set of polynomial dependences, including cross-terms among the
different indices. Because of the extensive list of terms, some of which 
involve quadratics, the coefficients of each term can be large. 
The result is that the final value of [Fe/H] for a star is often obtained 
by taking differences among comparably large terms, the net sum of
the small differentials
resulting in an [Fe/H] value between +0.5 and --3.0. It must be emphasized that
this approach should lead to a calibration as good as if not better than
that developed by the simpler, traditional differential technique, and the
tests of the final calibration by \citet{SN89} seem to bear that out. 
>From the standpoint of a user, the one weakness of the calibration is
the difficulty in deciding the degree to which variations among the
different stellar parameters - temperature, surface gravity, and metallicity -
influence the final photometric abundance through the various indices.
Disentangling the effects of the various terms in the calibration will
be the focus of next section.

\section{The Source of the Anomaly}
If there are weaknesses with the G-star calibration, one might assume
that the need exists for a full-scale recalibration of the 
metallicity function over the entire
cool star range at all [Fe/H]. Though clearly desirable, it isn't 
necessary for the purposes of this investigation and, in fact, is 
not be feasible with the spectroscopic sample available today. From 
the cluster data discussed above,
the field star data of Fig. 2, and tests of the calibration from
spectroscopic comparisons \citep[e.g.,][]{E93}, with minor
modifications discussed below, the relation appears to
work reasonably well for the hotter stars, i.e., $b-y$ less than 0.47, at all
[Fe/H], though there is a modest color dependence to the zero point
of the abundances even among the hotter stars. The issue is the 
source of the discrepancy among the stars in
the cooler half of the calibration range, the late
G and early K type stars with [Fe/H] approximately solar or higher.  
Despite a number of investigations of dwarf stars over the last decade, 
the number of cooler G and hotter K dwarfs of approximately solar 
abundance with spectroscopic abundance determinations remains scant; most 
surveys have concentrated on stars of lower metallicity within the 
old disk and/or halo, as did the original photometric study by \citet{SN89}.
Moreover, in most spectroscopic studies, the number of stars 
surveyed remains small, leading to difficulties in ensuring that 
the abundances obtained from one analysis are on the same system as 
that of another. For this investigation, the best, internally consistent, 
spectroscopic sample of significant size for which $uvby$ data are 
available comes from \citet{FA97}.   

Following the approach adopted in many of the recent studies,
the 91 stars of \citet{FA97} were checked against the $uvby$ catalog of
\citet{HM98}, providing photometry for 35 stars between $b-y$ = 0.38 and 0.60.
This photometry was then adjusted to the system of \citet{SN88} using the
offsets of \citet{OLS93} since the majority of the composite photometry is based
upon the work of \citet{OLS83, OLS84, OLS93, OLS94}. The sample of 35
stars was divided into 2 groups with a boundary at $b-y$ = 0.47; 23 stars
fell within the hotter regime while 12 defined the cool subset.  The
photometric abundances are plotted as a function of the spectroscopic values
in Fig. 3, where the hotter stars are open squares and the cool stars are
filled. The difference in slope between the two data samples is obvious.
A linear fit for the hotter stars produces the relation $\rm{[Fe/H]}_{phot}$ =
0.85 $\rm{[Fe/H]}_{spec}$ + 0.02, identical within the uncertainties with the 
trends found by \citet{AL96} and \citet{HA01}. In contrast, the 12 
cooler dwarfs define a relation 
$\rm{[Fe/H]}_{phot}$ = 0.39 $\rm{[Fe/H]}_{spec}$ -- 0.19, with 
very small scatter. If the small sample of cooler dwarfs is representative of 
the general population, the sharp changes seen in the distributions of 
Figs. 1 and 2 are not merely the product of a flaw in the zero-point of the 
metallicity calibration with increasing color, but are due in large 
part to a serious error in the slope.

If there is a significant problem with the G-star calibration that goes beyond
an issue with the zero point, the source of the error remains hidden within
the functional form of the metallicity calibration. To gain some insight
into which of the variables may hold the key, we plot in Fig. 4 the correlation
of spectroscopic [Fe/H] (open stars) and photometric [Fe/H] (filled squares) 
with $m_1$ for the 12 cooler stars in Fig. 3. Other than the predictable 
fact that the photometric abundances cover a more modest range in [Fe/H], 
there is no significant difference between the patterns exhibited by the 
data as a function of $m_1$. There is also little correlation between [Fe/H] 
and $m_1$, implying that, other than increasing the scatter, random 
variations in $m_1$ are unlikely to propagate through the G-star 
calibration to produce the correlated deviations with [Fe/H] seen in Fig. 3. 
In Fig. 5, we repeat the same exercise for $c_1$. The difference is striking. 
For both abundance estimates, increasing [Fe/H] is highly correlated with 
increased $c_1$. In fact, for the
photometric values, $c_1$ alone provides an excellent indicator of [Fe/H]. We
emphasize that this trend is not an artifact of the small sample. We have
repeated the exercise for the $uvby$ catalogs of \citet{OLS93,OLS94} and find
the same result for those samples, reflecting the result that the G-star 
calibration has a strong dependence on $c_1$, in contrast with the results for
hotter dwarfs in the field \citep{CR75, SN89,E93} or in clusters
with turnoffs in the F star range \citep{TW83,NTC,ATT87}.  
What Fig. 5 indicates is that the size
of the dependence of [Fe/H] on changes in $c_1$ for the cooler star
metallicity calibration is too small. If the spectroscopic abundances of
Fig. 5 and the linear relations of Fig. 3 are correct, for cool dwarfs near
solar abundance the variation of [Fe/H] with $c_1$ should be 
approximately doubled.

\section{The Impact}
Before discussing the specific role of the calibration error in solar
neighborhood stellar analyses, a more general issue is the relevance of the G-star
calibration within such studies. As noted earlier, the original intent of
the \citet{SN89} paper was heavily weighted toward stars of lower metallicity,
and the long term history of the application of the F and G dwarf calibrations
bears out this view. The use of the calibration for stars of solar metallicity
would appear to be pushing the limits of the viability of the technique.
Errors should be expected and have been noted in every study that has made
extensive use of the calibration. The approach taken in refining the G-star
calibration to adjust for these deficiencies has been virtually identical in
every case and mimics the procedure used to revise the F-star calibration
by \citet{E93}, as illustrated by the technique used in the construction
of Fig. 3. Abundances are derived for stars using the original 
G-star calibration and compared with the updated spectroscopic data 
for the same stars. A linear transformation is derived that permits the 
photometric abundances to be translated to the spectroscopic system.

Such internal checks have formed the appropriate basis justifying the use
of the original G-star calibration in every major study of the last two
years \citep{RP00,LA00,HA01,FE01}, and each study has corroborated the
conclusion of the earlier work. With only a minor modification to the
slope and/or zero point, the original G-dwarf metallicity calibration
provides reliable relative metallicity estimates for stars over the
entire color range and metallicity range. For stars near Hyades metallicity,
the repeatedly calculated offset lies between 0.0 and +0.1 dex.

What Figs. 1 and 2, in conjunction with the analysis in Sec. 3, demonstrate 
is that no simple linear transformation or zero-point adjustment is 
capable of placing the stars within the G-dwarf range on an internally or 
externally correct metallicity scale. Moreover, the minor adjustments 
derived to date to modify the photometric abundances are totally inadequate. 
Though the hotter stars require offsets between 0.0 and +0.1 dex, the cooler
half of the color range demands adjustments between +0.1 and +0.4 dex,
the size of the offset being strongly dependent upon the color of the
star. As with the original calibration, the failure to identify this
problem lies with the dominance of the hotter dwarfs in the spectroscopic
comparisons used to test and revise the calibration.

If the primary flaw in the G-star calibration lies with the $c_1$ dependence,
how will this impact studies based in part upon a sample of such stars? 
Though the abundances of cooler stars will be systematically in error, 
the key question is whether or not the problem will affect all stars 
with equal severity. Unfortunately,
the answer is no. The $c_1$ index for cooler stars has long been used as
a means of separating dwarfs from giants and subgiants 
\citep{OLS84,PSB,ATT94}, but these samples have involved stars of 
primarily halo and/or mixed [Fe/H]. To see the degree to which changes 
in luminosity alter
the $c_1$ index at a fixed [Fe/H] comparable to the metallicity of the disk,
we turn to the discussion of \citet{TAB}. This investigation
includes an analysis of the distance to the LMC using the open
cluster NGC 2420 as a link between the main sequence of the 
theoretical isochrones, as fixed by the solar neighborhood dwarfs \citep{TAB}, 
and the LMC. The field stars with {\it Hipparcos} parallaxes were 
selected using the metallicity provided by the 
G-dwarf calibration under discussion and required to lie between [Fe/H] =
--0.3 and --0.5, with an assumed cluster [Fe/H] of --0.4. By pure luck, the
one region where the G-dwarf calibration overlaps with the spectroscopic in
Fig. 3 is at [Fe/H] = --0.4, so the stars under consideration did, in the
mean, have approximately the same [Fe/H] of --0.4. For the stars near $b-y$ =
0.5, the $c_1$, $b-y$ diagram allowed perfect separation of subgiants at
the base of the giant branch from 
unevolved main sequence stars of the same color. A difference of 0.06 in
$c_1$ translated into a difference of over two magnitudes in $M_V$, a less
dramatic effect by a factor of two to three than found among halo 
stars \citep{PSB, ATT94}, but readily
detectable among disk field stars with precision photometry. Differences in
$c_1$ significantly larger than this should imply a more evolved star, i.e.,
a giant, at a given color near solar [Fe/H]. 

Among cooler stars, two factors can generate significant changes in $c_1$. For
unevolved stars at a given $b-y$, higher [Fe/H] leads to larger $c_1$. Second,
at a given $b-y$, luminosity alters $c_1$ in that more evolved stars have 
larger $c_1$. The effect is not limited to subgiant versus dwarf; the $c_1$
index should also change as a star evolves away from the zero-age-main-sequence
(ZAMS) with age. Though the latter effect is smaller, a modest evolutionary
effect can mimic that of a metallicity change, and therein lies the problem.
The weakness of the G-dwarf calibration is that it does not adjust the 
metallicity enough to account for the larger value of $c_1$ expected because
of the higher [Fe/H]. However, if a
star is significantly evolved away from the ZAMS, but still not a subgiant,
the increased $c_1$ due to evolution could compensate for the inadequate
size of the term in the calibration. Thus, one might expect that 
some metal-rich stars will be correctly identified as such as long 
as they are sufficiently evolved.

To test this possibility, we have run the catalogs of \citet{OLS93,OLS94} 
through the G-star calibration and selected those stars which 
have $\rm{[Fe/H]}_{phot}$ greater than
0.10 and $b-y$ greater than 0.45, a horizontal cut through the Figs. 2a and
2b. These stars were matched with the {\it Hipparcos} parallax catalog 
\citep{PE97}. Any star with $\sigma_{\pi}$/$\pi$ less than 0.15 was adjusted
for the Lutz-Kelker correction \citep{LK73,KO92} and plotted on a CMD. It was
assumed that the stars had no reddening. The result is shown in Fig. 6. Also 
shown is the main sequence for the Hyades using the composite sample 
constructed by \citet{ATTY} and the absolute magnitudes of \citet{PE98}.
Stars that exhibit any indication of multiplicity from the summary by
\citet{deb} have been excluded.

The single-star main sequence (filled circles) of the cluster 
is extremely tightly defined. In contrast, the field stars from \citet{OLS93}
(open triangles) and \citet{OLS94} (squares) extend over a wide range in 
absolute magnitude with a remarkable cutoff at $b-y$ = 0.485. The number
of stars between $b-y$ = 0.49 and 0.57 is small, with a modest number 
reappearing at $b-y$ = 0.58. This peculiar distribution is a reflection
of the calibration problem noted in Fig. 1, coupled with the selection
biases of the samples and the red limit of the subgiant branch. Stars 
that are Hyades metallicity or higher have photometric abundances 
that are systematically underestimated
among the cool stars. Only at the red limit of the G-star metallicity
calibration ($b-y$ $\sim$ 0.58) does this effect weaken due to the
curvature seen in Fig. 2. However, what is 
more important from the standpoint of this discussion is the 
nature of the stars that are picked as being metal-rich.

As one moves down the main sequence toward cooler temperatures, the field
star distribution reaches an approximate limit in $M_V$; few stars are
found fainter than $M_V$ = 5.5 until the cool limit of the sample is attained.
This implies that between $b-y$ = 0.45 and 0.57, a star must be located
in a CMD position increasingly above the unevolved/partially evolved
main sequence of the Hyades. It is unlikely that the stars in this color 
range in Fig. 6
are located above the main sequence because they are binaries. Composites
of two, almost identical, main sequence stars will 
shift the system about 0.7 mag
above the main sequence, but will not significantly alter the color indices.
Thus, a composite system will be affected by the photometric metallicity
error to the same extent as a single star. The stars located between
$b-y$ = 0.47 and 0.57 have survived the cut because of some 
combination of (a) very high metallicity, (b) evolution off the main sequence, 
and (c) photometric errors that scatter them into the high metallicity
regime. 

This point is clarified in Fig. 7, where the $c_1$, $b-y$ diagram for the stars
in Fig. 6 is presented; the filled circles are the Hyades data of 
\citet{OLS93, OLS94} transferred to the same system as the field stars. Note
that the scatter in Hyades $c_1$ is the primary source of the [Fe/H] scatter
in Fig. 1. It is apparent that the two-color diagram is a reflection of the
the CMD; the $M_V$ cut in Fig. 6 is translated into a diagonal cut in the
$c_1, b-y$ diagram, an effect which disappears at the cool limit of the
sample. What is surprising is the set of $c_1$ values for the seven stars 
between $b-y$ = 0.5 and 0.55. Despite the fact that these stars are
positioned only 0.2 to 0.7 mag above the Hyades relation in the CMD, implying
that they are not subgiants but main sequence stars, their $c_1$ indices
place them 0.08 to 0.15 mag above the Hyades relation, classifying 
them photometrically as giants. Since they are not significantly evolved and
the size of the difference in $c_1$ is too large to be caused by photometric
error, either these stars suffer from some form of atmospheric anomaly
that distorts their indices, or they are very metal rich.

The seven extreme stars are HD 7199, 24257, 77712, 87000, 100508, 122640, and
145675. Perhaps the most studied of these is HD 145675 (14 Her).  This 
star has been the focus of a good deal of attention
because of the discovery of planets around the system \citep{MA99}, as well as
confirmed, spectroscopic abundances that classify the star as super-metal-rich,
with [Fe/H] above +0.3 \citep{TA96,GWS}. Its virtual twin, HD 75732,
occupies the same areas of the CMD and the color-color diagram, but missed the 
cut in [Fe/H] because its photometric abundance is +0.09. For the remaining
six stars, no additional information is available which might cast some light
on their location in these figures. If HD 145675 is typical of this select
group, its location in the CMD well above the Hyades main sequence is a
consequence of some combination of the shift in the ZAMS to higher 
$M_V$ at a given $b-y$ due
to an increase in [Fe/H] of between 0.3 and 0.4 dex relative to the Hyades and
a greater degree of evolution. The increasing gap between the Hyades main
sequence and the field stars starting at $b-y$ = 0.47 is likely to be a
product of the growing discrepancy between the photometric and true abundance
toward redder color, requiring an even higher intrinsic [Fe/H] and/or greater
degree of evolution to meet the fixed cut in the photometric [Fe/H].  

The luminosity effect on $c_1$ and, indirectly, on [Fe/H], provides a
natural explanation for the bluer break point in the distribution for
the cluster stars compared with the field (see Figs. 1 and 2). 
The cluster stars at a given
color are in the same evolutionary state and only moderately metal rich.
The field star sample includes stars significantly more metal rich than
the Hyades, thus redder at a given temperature, and covering a range
in evolutionary state from the unevolved main sequence to subgiants at
the same color.

An approximate estimate of the significance of the calibration distortion
on a sample selected for purposes of stellar population analysis 
may be obtained in the following way. For the Hyades stars with $uvby$ 
photometry as plotted in Fig. 1, a mean photometric [Fe/H] has been
derived as a function of $b-y$ between 0.39 and 0.57. Assuming the 
error in the photometric abundance is purely a zero-point error in the 
calibration, an offset can be calculated as a function of $b-y$ to bring the
mean photometric abundance in line with that adopted for the cluster as
a whole, in this case, [Fe/H] = +0.125. This color-dependent offset is then 
applied to any star processed through the calibration. Since we have already
demonstrated that the error lies in a combination of both the 
slope and zero-point of the calibration (see Fig. 3),
this crude fix will only give a reasonable adjustment to stars more metal-rich
than the Hyades, with the uncertainty increasing for more metal-rich stars.

The resulting metallicity histogram for stars from the catalogs of
\citet{OLS93,OLS94} with photometric 
[Fe/H] above +0.2, higher than the adopted Hyades value of +0.125,
is shown in Fig. 8, where the solid line is the distribution based upon
the corrected [Fe/H] and the dashed line is the distribution based
upon the uncorrected G-star calibration of \citep{SN89}. Such a 
sample would be appropriate in a search for metal-rich dwarfs to
study as potential sources of planets, as exemplified by \citet{LA00}. We
emphasize that no attempt has been made to remove spectroscopically 
peculiar stars, binaries, or spectroscopic giants in either sample. Our
primary concern is the differential effect of the adjustment which clearly
expands the sample of very metal-rich stars in both number and range. 
Moreover, the boundary of [Fe/H] = +0.2 is arbitrary; one could lower this
to --0.1 and the pattern would remain the same. The
source of the enhancement is shown in Fig. 9, where the samples of Fig. 8
are sorted by color. Superposed upon the bias in the original catalogs,
which produces a decline in the number of stars redder than $b-y$ = 0.50, 
the metallicity adjustment leads to  a significant number of high-metallicity
stars at all colors, with the greatest enhancement at the red 
end of the distribution. Part of the decline in numbers at 
redder colors in both samples can also be attributed to the
cool limit of the subgiant branch at higher [Fe/H]. Stars at the base
of the giant branch will be selectively enhanced in number because they
can be observed over a greater range in distance and because of potential
effects on $c_1$ caused by evolution rather than metallicity.

\section{Summary and Conclusions}
Field star $uvby$ data from the high precision, photometric catalogs of 
\citet{OLS93,OLS94} and cluster photometry for the Hyades and Praesepe
have been analyzed to test potential issues involving the G-star metallicity
calibration of \citet{SN89}. The reliability of the photometric approach
has taken on increasing importance in bridging the gap between the extensive
sample of stars with reliable kinematic and distance information and
the small to modest number with consistent abundance estimates.

The field star and cluster data tell the same story: for dwarf stars of 
mid-G to early K, the metallicity calibration systematically underestimates
the metallicity of the more metal-rich stars in the solar neighborhood, 
with the
size of the discrepancy increasing toward higher [Fe/H]. Among the hotter half
of the sample, the linear relation between photometric and spectroscopic 
abundances is similar to what has been found in earlier 
analyses \citep{AL96,HA01}, but the cooler stars generate a linear 
relation that has half the slope
of the hotter stars (see Fig. 3). It must be kept in mind at 
all times that while the
reality of the discrepancy is obvious, the sample of stars that
define the change in the linear relations is small. It should not be
assumed that the problem can be corrected by simply subdividing stars into
two color groups. The required adjustments may depend upon the slope, the
zero-point, or both, and the relative contribution of either parameter may
vary with $b-y$.

The source of the problem appears to lie with the metallicity dependence
of the $c_1$ index. The reason for its long-standing survival in the 
calibration is best illustrated
by comparison with the $uvby$ metallicity calibration for F stars. The
most commonly adopted calibrations for the hotter dwarfs are those of
\citet{CR75} and \citet{SN89}, modified in a modest way by \citet{E93}.
The F-dwarfs were a focus of the early work on age determination among
solar neighborhood stars because they had the greatest potential for
exhibiting large deviations from the ZAMS in an age range comparable to
the lifetime of the disk, an important property given the need for precise
determination of absolute luminosity from photometric or trigonometric
methods. Thus, the sample of stars with available spectroscopic data in
this temperature regime grew in response to the need to define
the stellar abundances reliably, culminating in the 
extensive sample of \citet{E93}.
Other than modest adjustments introduced by changes in the
spectroscopic abundance scale, the F-star calibrations have remained quite
stable with excellent agreement among the various investigators. 
Repeated cross-checks with expanded samples have made errors of the type
noted above difficult to sustain.

The second, more critical factor has been the availability of cluster
photometry on the $uvby$ system. F-stars in a number of nearby open 
clusters were well within reach of traditional, photoelectric photometers
and the changeover to CCD's has expanded this option to even greater
distance. While permitting a check on the metallicity calibrations via
the chemically well-defined sample within a cluster, clusters of
intermediate and older age gave direct tests of the impact of main
sequence evolution on the metallicity calibration and the relationship
between $\delta$$c_1$ and $\delta$$M_V$. Repeated observations of
a large sample of clusters \citep[e.g.,][]{CB70,NTC,ATT87,SN88,DA94}
have confirmed the conclusion from field star data that the F-star
$m_1$ metallicity calibration has little to no dependence on $c_1$.
It must be emphasized that there is, at present, no evidence for any
systematic difference between spectroscopic and photometric $uvby$
abundances for F-stars in the field or in clusters beyond those required
by ongoing revisions in the spectroscopic scale.

In contrast, while the spectroscopic sample of G-dwarf abundances,
particularly at solar metallicity and above is growing, it still 
remains modest and not always internally consistent. Though
CCD photometry now places such observations within reach, the number of
open clusters where the cool G and K-dwarf members have been observed 
on the $uvby$
system remains limited to a few nearby open clusters \citep{REG91,RF92}.
The combination of these two factors is critical because the implication
of our analysis is that, contrary to what is found for the hotter stars,
the $c_1$ index for cooler dwarfs is strongly affected by variations
in [Fe/H]. The rapid rise in sensitivity occurs near $b-y$ = 0.47, making
more metal-rich dwarfs appear more evolved, i.e., have a higher $c_1$
index. The effect is so extreme that main sequence stars with metallicity
well above the Hyades value are photometrically classified as giants, a
secondary source of problems because the transformation between instrumental
and standard indices for cooler stars is often dependent upon the
luminosity classification of the star, as illustrated by the $Caby$
transformation in \citet{TAT95}. The absence of an extensive sample
of cool dwarfs with spectroscopically determined abundances at all
[Fe/H] covering varying degrees of evolution, particularly
at the metal-rich, cooler end of the distribution, makes adequate
determination of the impact of the metallicity effect on $c_1$ difficult,
if not impossible. The plausibility of the color effect illustrated
in Fig. 3 rests primarily upon its consistency with the unmistakable
distortions of Figs. 1 and 2.

The effect of the calibration error in the $uvby$ system on abundance
determination is straightforward, but caution is recommended in
interpreting its impact on stellar population studies. Cool dwarfs
of solar metallicity and above will have their abundances underestimated
by a significant amount and many metal-rich dwarfs will be incorrectly
excluded from study because they are photometrically classed as giants.
However, metal-rich stars that are well-evolved off the main sequence
or located in the subgiant region will see some partial correction to
their photometric abundances due to evolutionary effects on $c_1$. The
result could be that truly metal-rich evolved stars will be preferentially
selected in surveys for metal-rich stars, while their unevolved 
counterparts are excluded.  Since the evolutionary effect only partially
corrects for the photometric underabundance, stars with parallaxes will
find themselves positioned even farther above a ZAMS defined by their
lower, photometric [Fe/H], leading to an excessively large
age determination for stars already selected because they are, on average,
older. Until a reliable means of disentangling the influence of luminosity
and metallicity effects on $c_1$ indices is obtained, photometric
$uvby$ abundances for cooler field stars, and stellar population
analyses tied to them, should be regarded with skepticism.

\acknowledgements
The authors gratefully acknowledge the thoughtful comments and 
suggestions of the referee, Dr. Ken Yoss, which made the publication
of this paper possible.

\clearpage
\figcaption[fig1.ps]{The metallicity of the Hyades (filled symbols) and
Praesepe (open symbols) as a function of color using the $uvby$ G-star
calibration of \citet{SN89}. \label{fig1}}

\figcaption[fig2.ps]{The metallicity distribution as a function of
$b-y$ for stars in the catalog of \citet{OLS93} (a) and in \citet{OLS94}
(b) using the G-star calibration of \citet{SN89}. \label{fig2}}

\figcaption[fig3.ps]{Comparison of photometric and spectroscopic abundances
for stars with $b-y$ less than 0.48 (open symbols) and those redder
than $b-y$ = 0.48 (filled symbols). Dashed relation shows the linear fit
to the hotter stars, the solid line the relation for the cooler stars.
\label{fig3}}

\figcaption[fig4.ps]{Spectroscopic (open stars) and photometric (filled squares)
abundances as a function of
$m_1$ for the 12 cool stars of Fig. 3. \label{fig4}}  

\figcaption[fig5.ps]{Same as Fig. 4 versus $c_1$. \label{fig5}}

\figcaption[fig6.ps]{CMD for field stars with reliable parallaxes
and photometric abundances greater than +0.10. Open squares are
stars from \citet{OLS93}, open triangles are from \citet{OLS94},
and filled circles are Hyades stars. \label{fig6}}

\figcaption[fig7.ps]{Color-color plot for same stars in Fig. 6. \label{fig7}}

\figcaption[fig8.ps]{Distribution of photometric abundances for stars
from \citet{OLS93,OLS94} between $b-y$ = 0.39 and 0.57 with [Fe/H] 
above +0.2. Dashed line 
refers to the original $uvby$ calibration while the solid curve includes
the adjusted abundances. \label{fig8}}
 
\figcaption[fig9.ps]{Distribution of abundances as a function of
color for the same samples found in Fig. 8.
\label{fig9}}
\enddocument
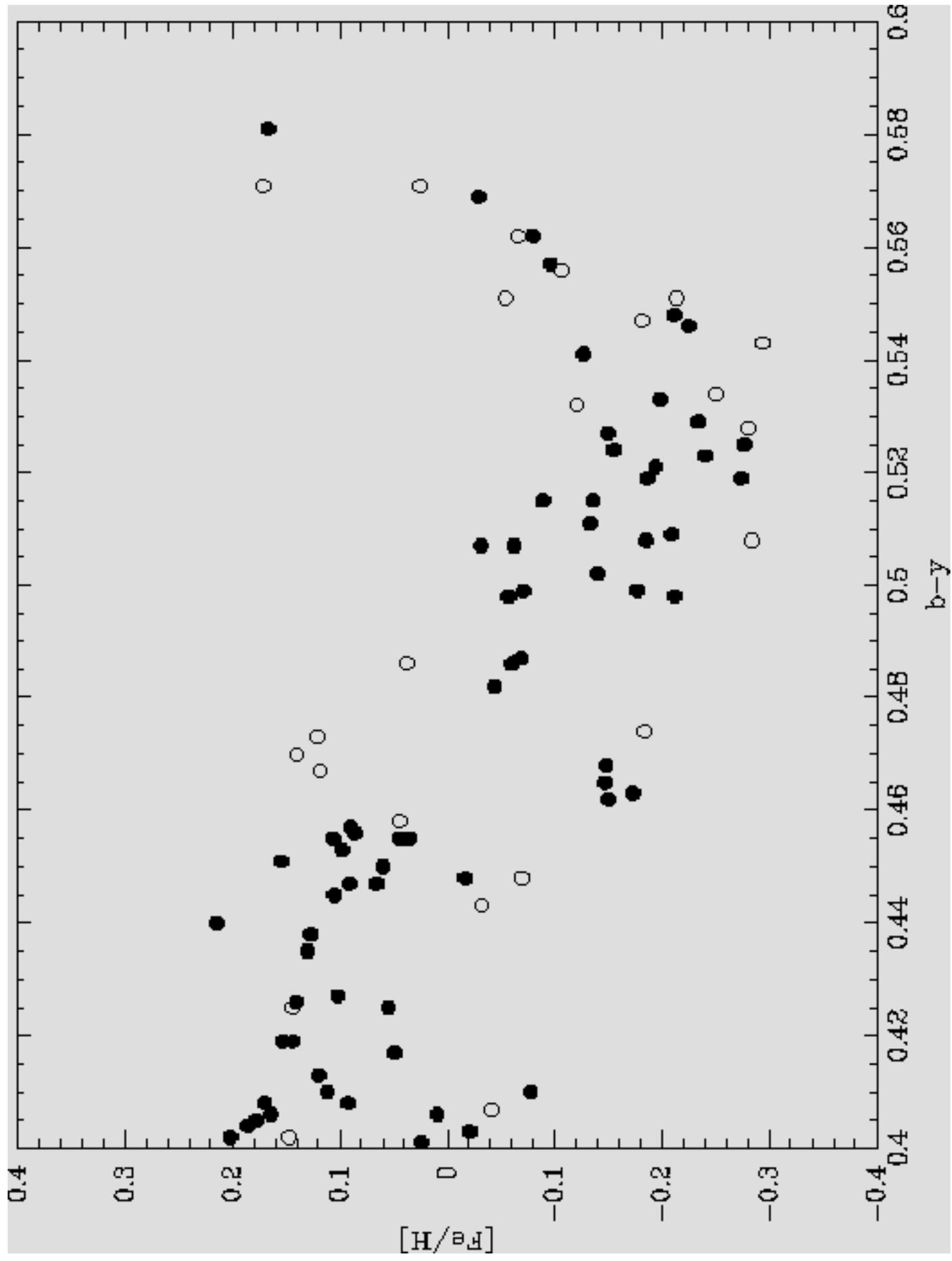
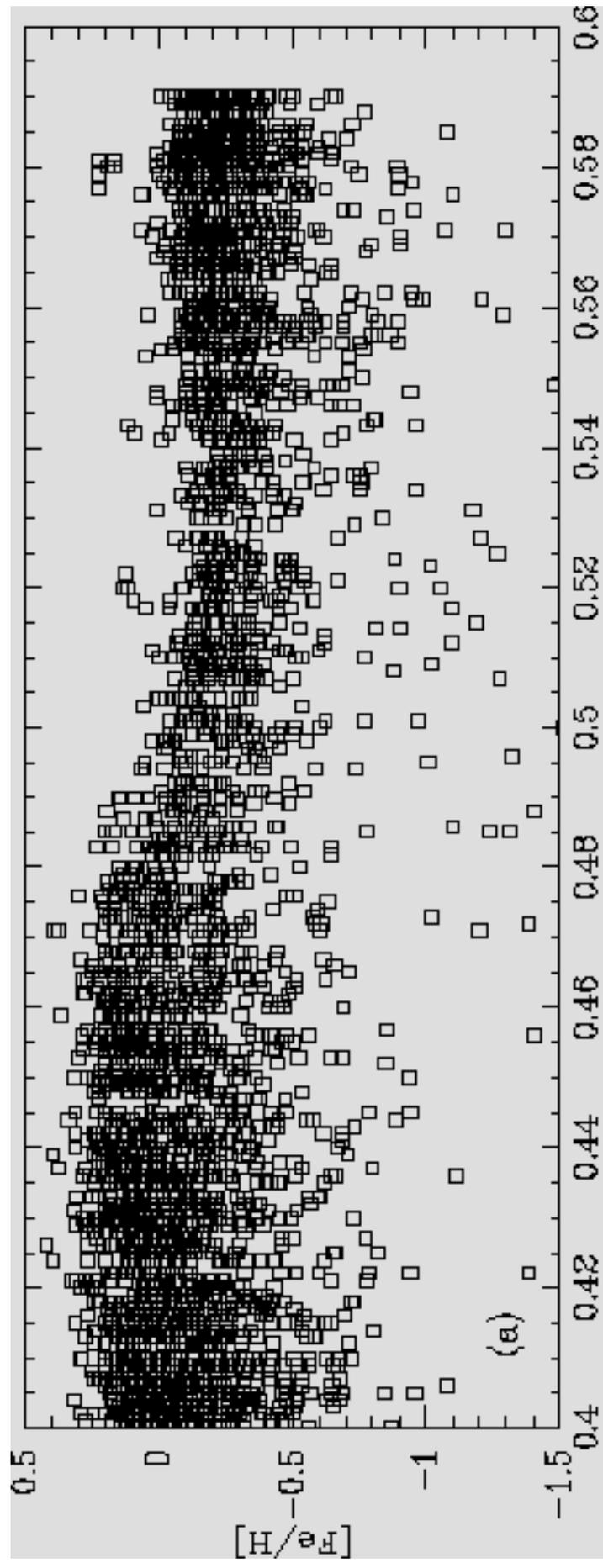
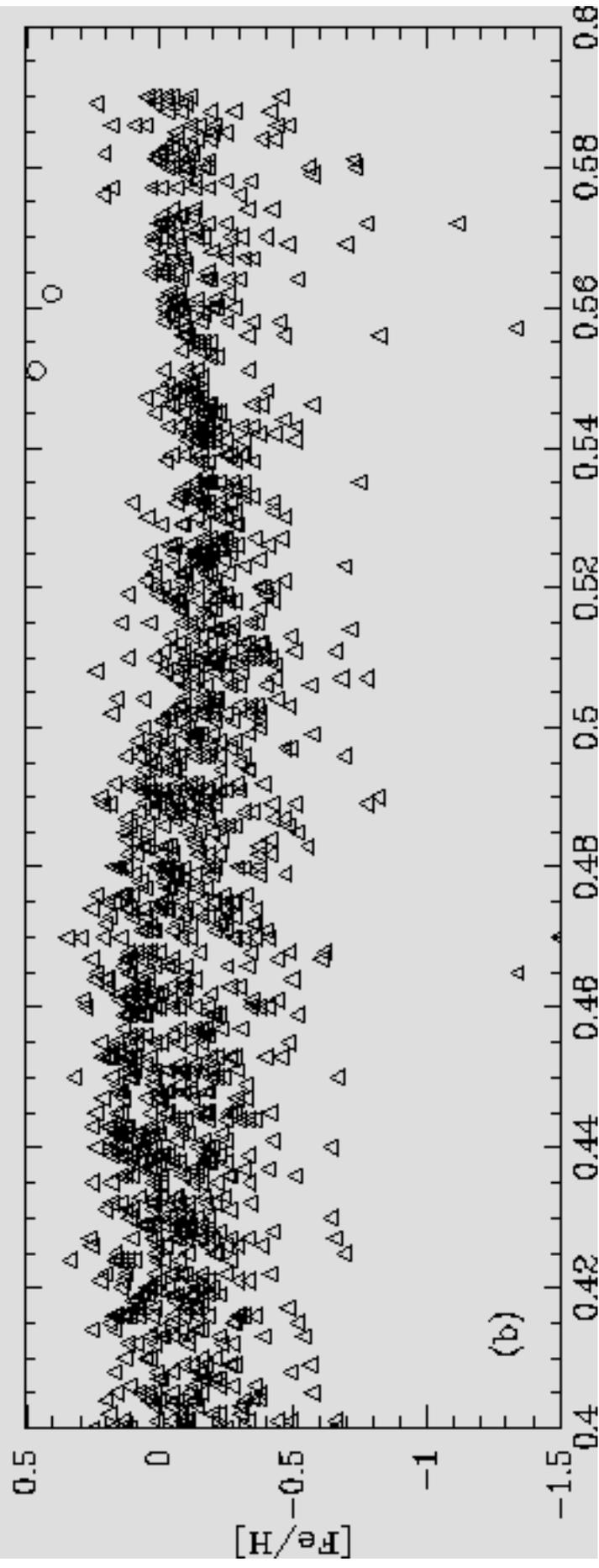
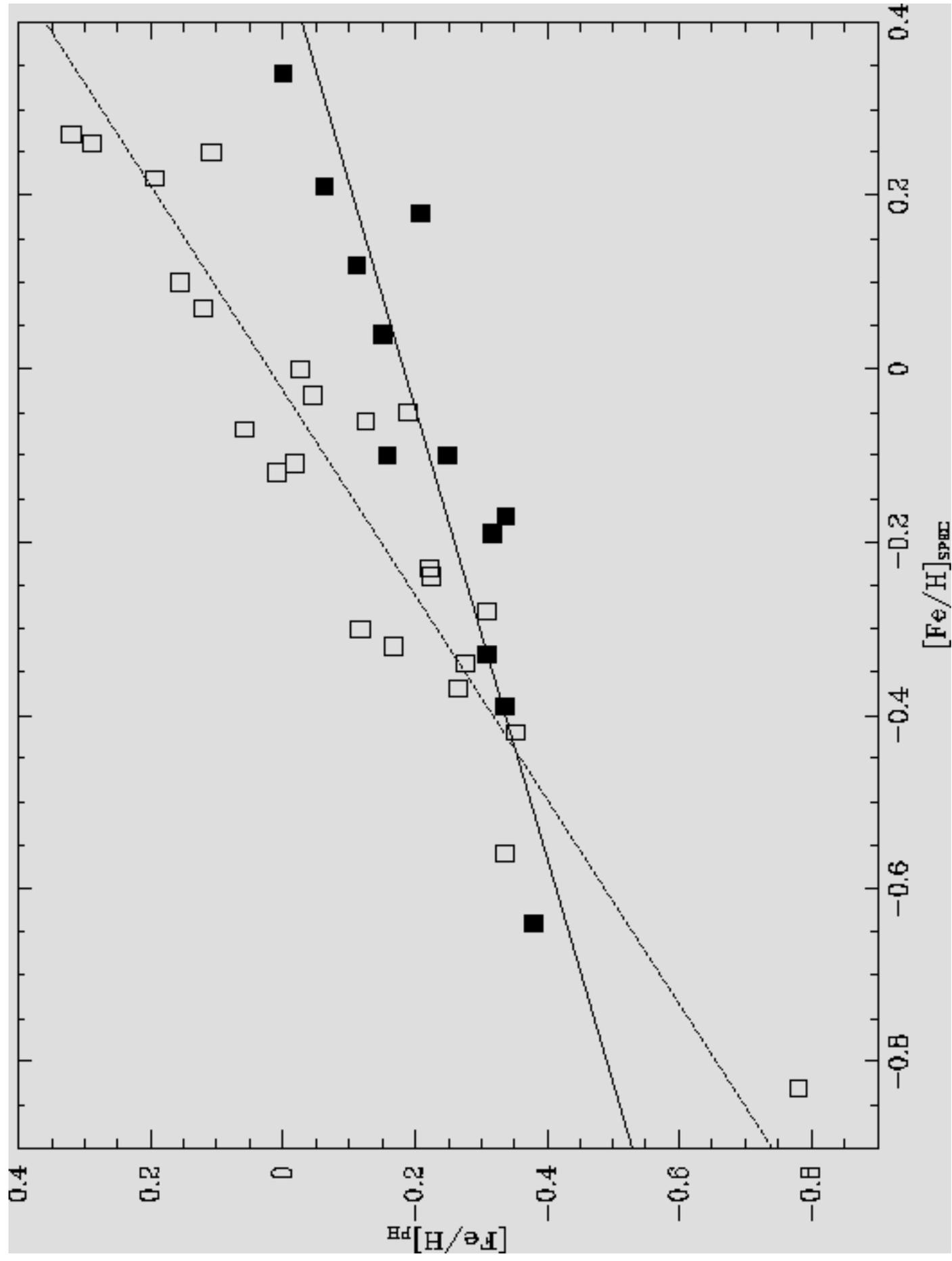
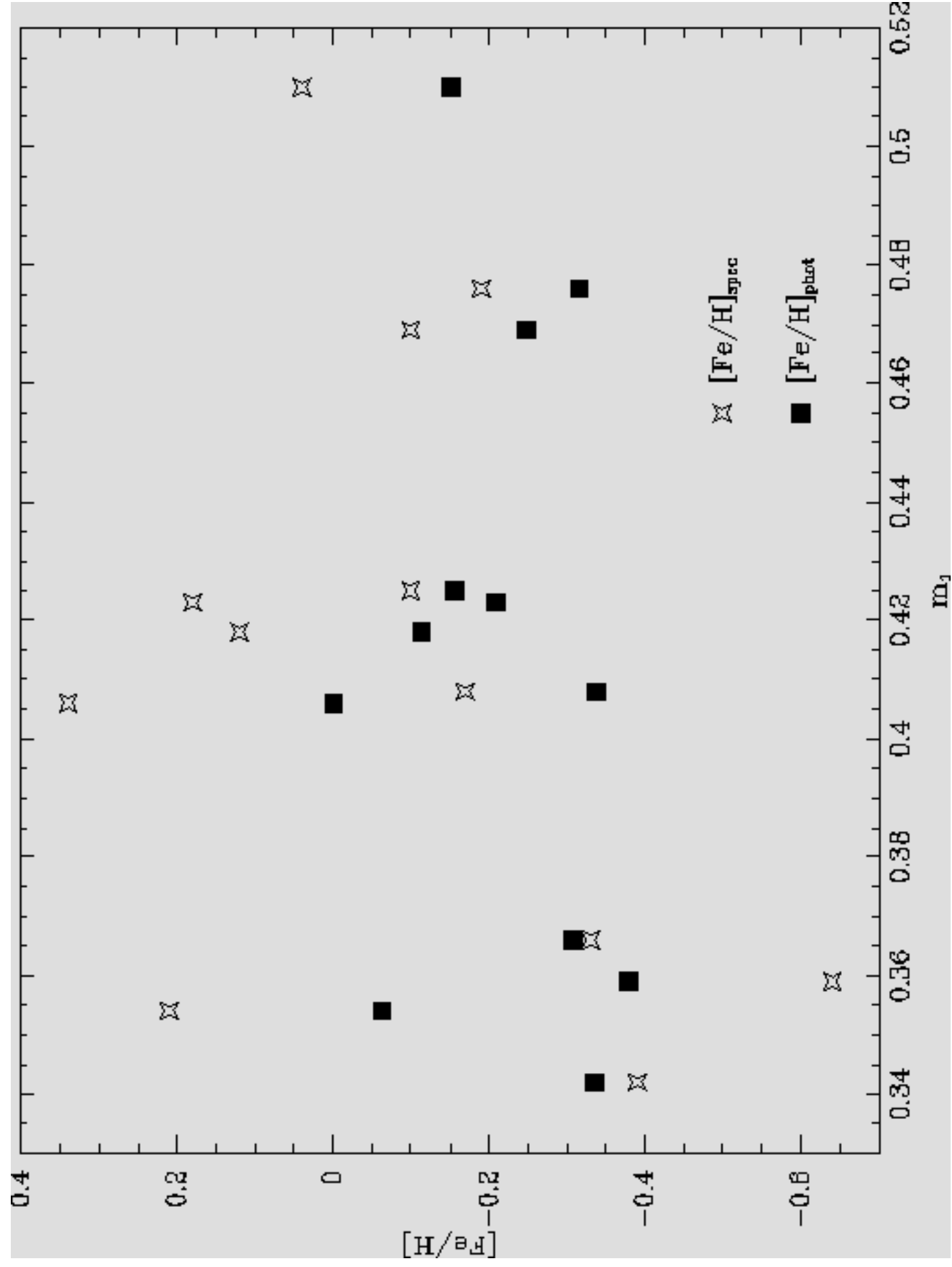
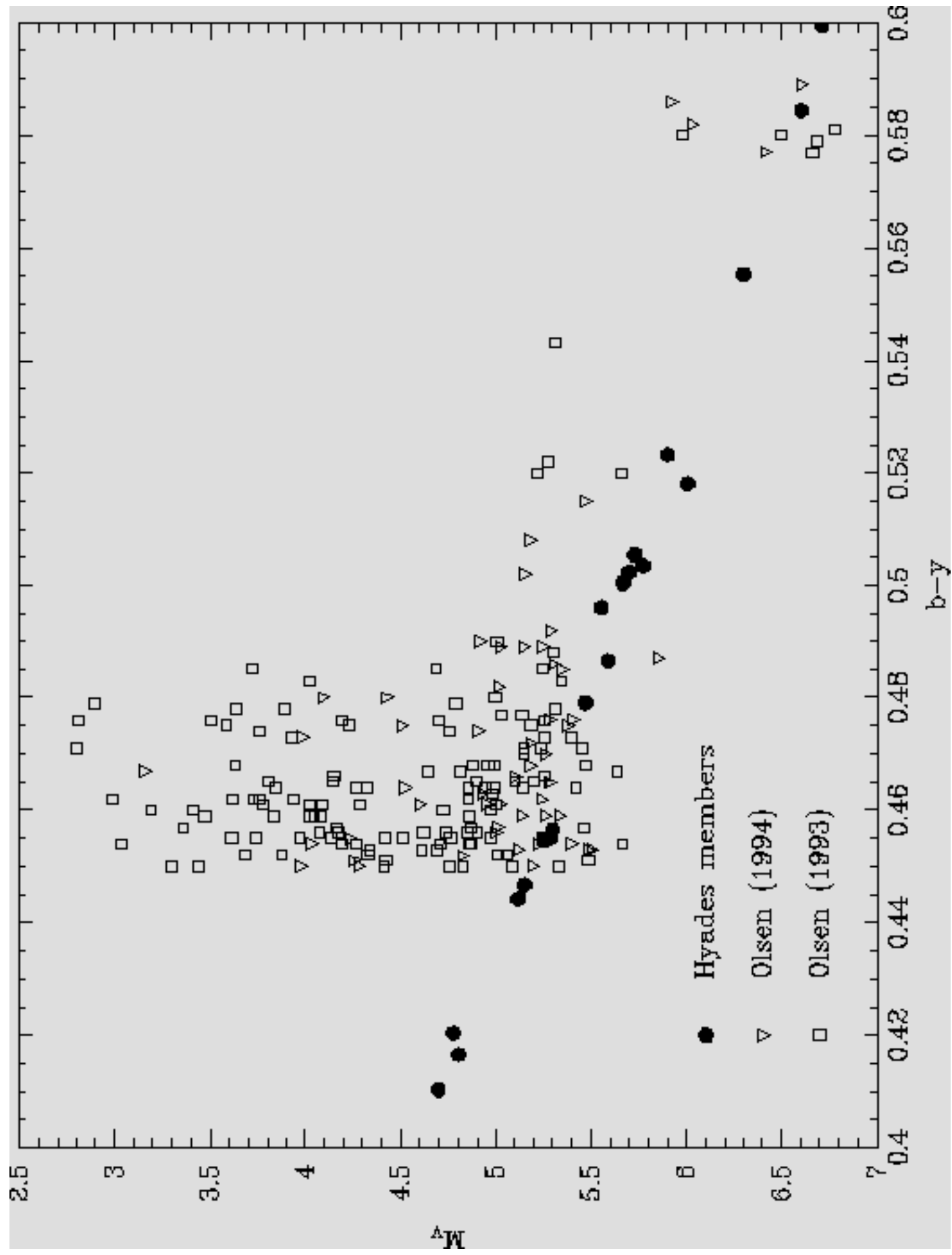
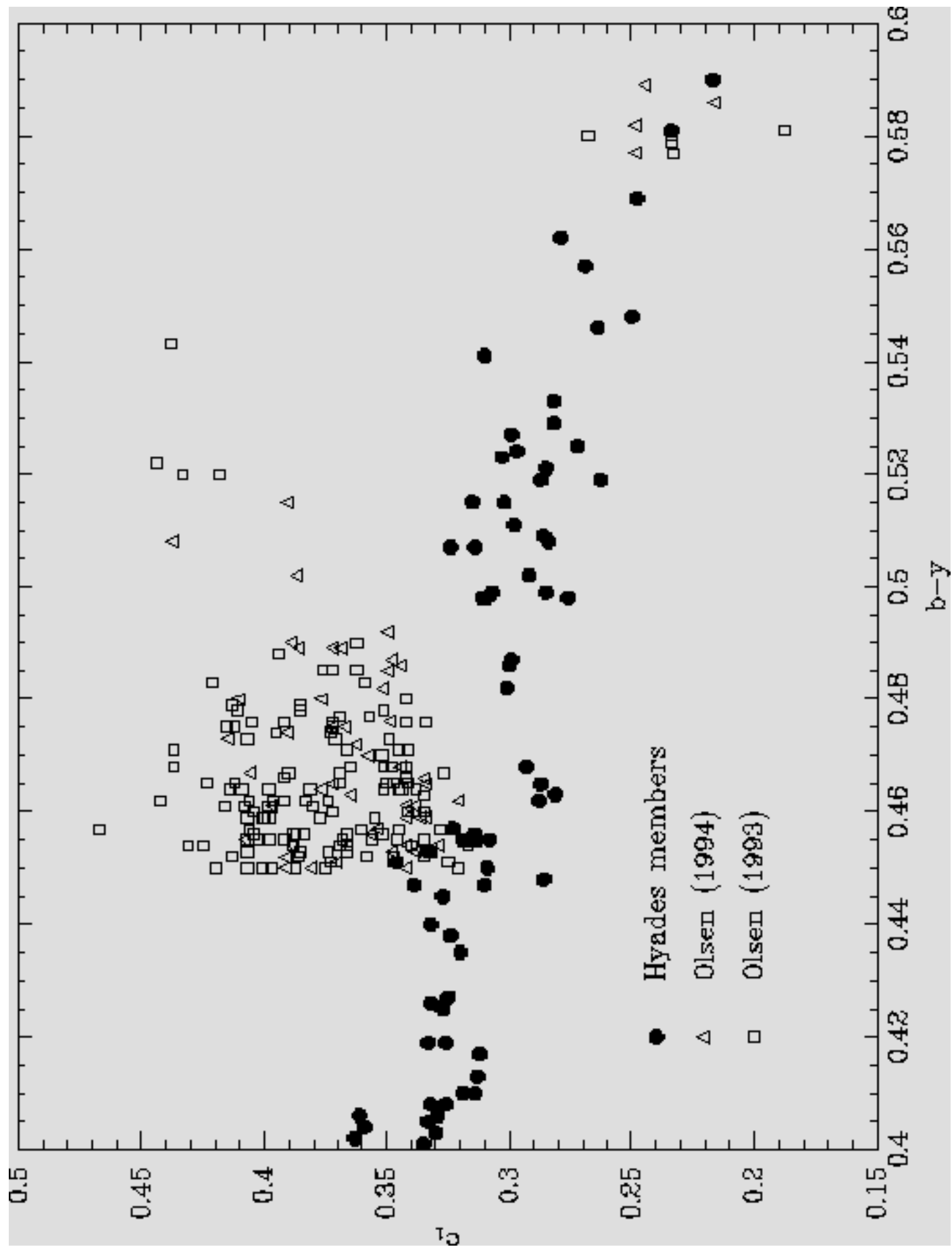
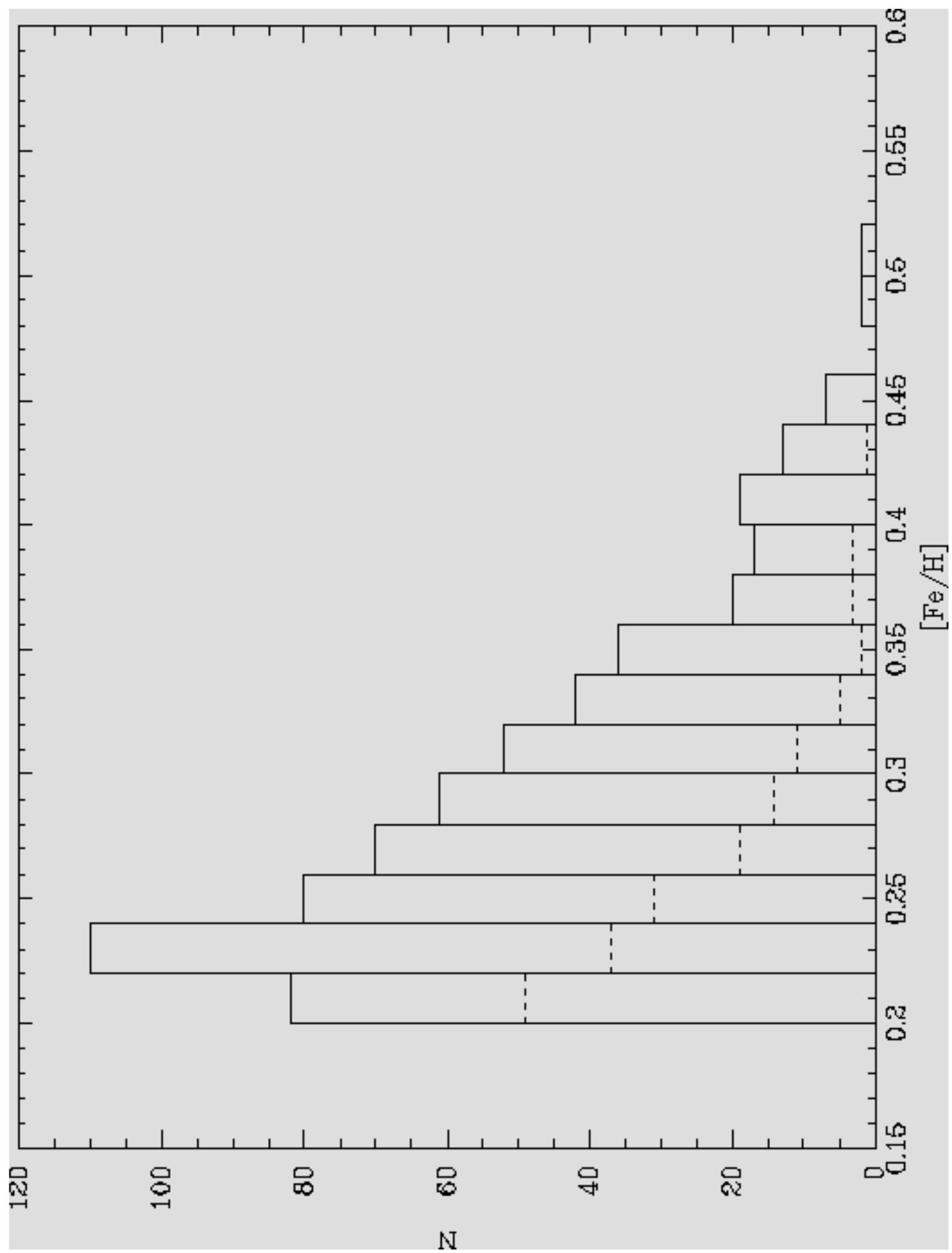
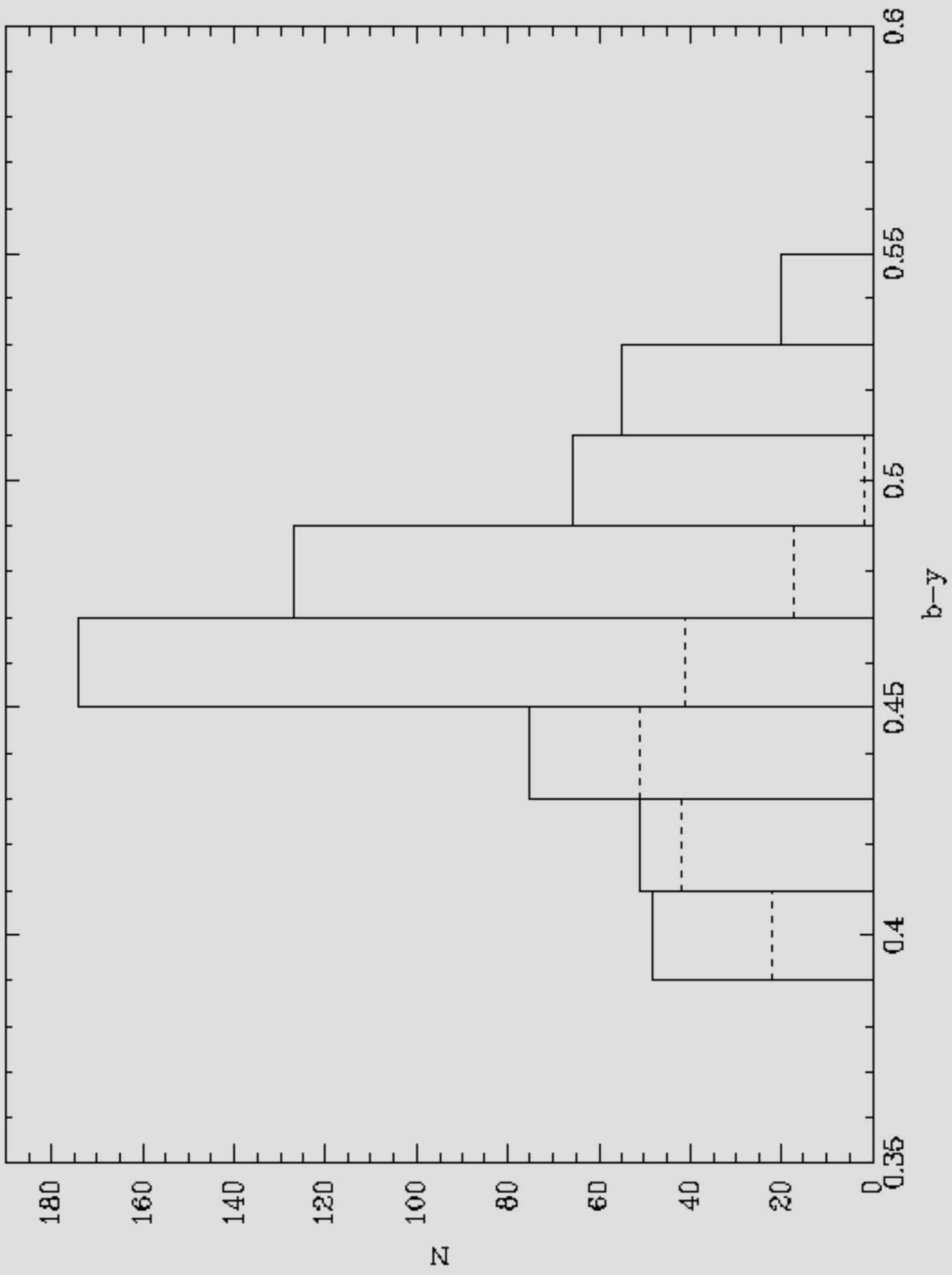
